\RequirePackage{amsmath} %
\documentclass[runningheads]{llncs}

\usepackage[utf8]{inputenc}
\usepackage[T1]{fontenc}
\usepackage[font=scriptsize, farskip=5pt]{subfig}
\usepackage{graphicx}
\usepackage{url}
\usepackage{xcolor}
\usepackage{wrapfig}
\usepackage{algorithmicx}

\usepackage{tabularx}
\newcolumntype{C}[1]{>{\centering\arraybackslash}p{#1}}
\newcolumntype{R}[1]{>{\raggedleft\arraybackslash}p{#1}}
\newcolumntype{L}[1]{>{\raggedright\arraybackslash}p{#1}}

\usepackage{booktabs}
\usepackage{amssymb}
\usepackage{amsmath}
\usepackage{siunitx}
\usepackage{microtype}
\usepackage{xparse}
\usepackage{xspace}
\usepackage{csquotes}

\usepackage{siunitx}
\usepackage{todonotes}

\usepackage{tikz}
\usetikzlibrary{calc, arrows, shapes, decorations.pathreplacing}

\definecolor{darkgreen}{rgb}{0,0.5,0} %

\definecolor{mylinkcolor}{rgb}{0,0,.5} %
\definecolor{mycitecolor}{rgb}{0,0.5,0} %
\usepackage[
colorlinks=true,
breaklinks=true,
linkcolor=mylinkcolor,
menucolor=mylinkcolor,
urlcolor=mylinkcolor,
citecolor=mycitecolor, %
]{hyperref}
\usepackage[capitalize]{cleveref} %

\newcommand{\etal}{\emph{et al.}\xspace}

\newcommand{\capsacronym}[1]{{\footnotesize \mbox{#1}}\xspace}
\newcommand{\capsacronyms}[1]{{\footnotesize \mbox{#1}}s\xspace}
\newcommand{\LSTM}{\capsacronym{LSTM}}
\newcommand{\LSTMs}{\capsacronyms{LSTM}}
\newcommand{\TCN}{\capsacronym{TCN}}
\newcommand{\TCNs}{\capsacronyms{TCN}}
\newcommand{\EPS}{\capsacronym{EPS}}
\newcommand{\MSE}{\capsacronym{MSE}}

\definecolor{cNorTb}{rgb}{1,1,1}
\definecolor{cNorTf}{rgb}{0,0,0}

\newcommand{\ltfmin}[1]{\underline{#1}}
\newcommand{\ltfmax}[1]{\fontseries{b}\selectfont #1}
\newcommand{\ltfbetter}[1]{\textcolor{darkgreen}{#1}}
\newcommand{\ltfworse}[1]{\textcolor{red}{#1}}

\newcommand{\var}[1]{#1\xspace}

\newcommand{\predAbbildung}			{\var{\ensuremath{\phi}}} %
\newcommand{\predFeaturewindow}		{\var{\ensuremath{\beta}}} %
\newcommand{\predHorizont}			{\var{\ensuremath{\tau}}} %
\newcommand{\predData}				{\var{\ensuremath{q}}} %

\newcommand{\dataset}[1]{\textsc{#1}\xspace}
\newcommand{\computstatquarterly}{\dataset{Computstat Quarterly}}
\newcommand{\dailyshares}{\dataset{Daily Shares}}

\newcommand{\dsparam}[1]{\texttt{#1}}

\begin{document}

\title{Earnings Prediction with Deep Learning}%
\author{%
	Lars Elend\inst{1}\orcidID{0000-0001-6256-7174}
	\and Sebastian A. Tideman\inst{2}\orcidID{0000-0003-1735-457X}
	\and Kerstin Lopatta\inst{2}\orcidID{0000-0003-0267-0588}
	\and Oliver Kramer\inst{1}\orcidID{0000-0001-7607-1700}
}
\authorrunning{L. Elend et al.}
\institute{%
	Computational Intelligence Group, Department of Computer Science,\\
	Carl von Ossietzky University of Oldenburg, 26111 Oldenburg, Germany\\
	\email{<firstname>.<lastname>@uni-oldenburg.de}
	\and
	Chair of Accounting, Auditing and Sustainability,\\
	University of Hamburg, 20146 Hamburg, Germany\\
	\email{<firstname>.<lastname>@uni-hamburg.de}
}
\maketitle              %
\begin{abstract}
In the financial sector, a reliable forecast the future financial performance of a company is of great importance for investors' investment decisions. In this paper we compare long-term short-term memory (\LSTM) networks to temporal convolution network (\TCNs) in the prediction of future earnings per share (\EPS).
The experimental analysis is based on quarterly financial reporting data and daily stock market returns.
For a broad sample of US firms, we find that both \LSTMs outperform the naive persistent model with up to 30.0\% more accurate predictions, while \TCNs achieve and an improvement of 30.8\%.
Both types of networks are at least as accurate as analysts and exceed them by up to 12.2\% (\LSTM) and 13.2\% (\TCN).

\end{abstract}

\section{Introduction}

Investors rely first and foremost on earnings predictions when making investment decisions, e.g., buy, hold, or sell a firm's shares.
Besides using own projections, they heavily rely on earnings forecasts provided by financial analysts. 
Consequently, forecasting earnings is one of the main tasks of financial analysts working at major financial institutions, e.g., broker firms. 
Analysts invest significant resources to provide accurate forecasts.
However, forecasting is a difficult undertaking as numerous factors have an influence on the prediction performance.
In this paper, we predict publicly listed US firms' quarterly earnings per share with state-of-the-art techniques from the field of deep neural networks based on companies' time series data.

We structure the remainder of this paper as follows. In \cref{sec:related}, we present related work on prediction of financial data.
The base time series model and quality measures are introduced in \cref{sec:tsm}.
We describe the data preprocessing process in \cref{sec:dp}.
Objective of our work is to compare \LSTM networks with \TCNs, which will be introduced in \cref{sec:deep}.
\Cref{sec:exp} presents the experimental analysis, and \cref{sec:cons} draws conclusions.

\section{Related Work}
\label{sec:related}

Analyst forecasts are often used to benchmark the accuracy of earnings predictions obtained from models.
However, due to recent regulation on financial analysts working conditions, e.g., limiting the private access to management, a drop in analyst coverage has been observed \cite{Anantharaman2011Cover}.
Automated earnings prediction models supported by artificial intelligence may fill this gap.
While there is already significant work on predictions of stock market price and returns (which is an aggregate of several factors such as firm-, industry-, country-level variables) using neural networks \cite{dosSantosPinheiro2017Stock}, empirical evidence is missing whether artificial intelligence can provide meaningful earnings forecasts as a direct measure of firm success.

Some evidence exists that fraud, e.g., illegal manipulation of earnings, can be predicted using machine learning \cite{Bao2020Detecting}.
In their study, Bao \etal (2020) find that ensemble learning with raw accounting numbers has predictive power for future fraud cases. Their approach outperforms logistic regression models based on financial ratios commonly used by prior research \cite{Dechow2011Predicting} as well as a support-vector-machine model \cite{Cecchini2010Detecting}, where a financial kernel maps raw accounting numbers into a set of financial ratios.
Yet, the prediction of restatements is relatively less challenging as it is a binary decision tree (future restatement vs. no future restatement). 
To the contrary, predicting future earnings is more challenging as all discrete values are theoretically possible and information from multiple sources, e.g., financial statements, stock market data, have to be considered.

To our knowledge, no study has yet predicted future earnings using artificial intelligence.
Closest to this study is the work of Ball and Ghysels (2018) \cite{Ball2017Automated}.
They use a mixed data sampling regression method (but no neural networks) to predict future earnings and find that their predictions beat analysts' predictions in certain cases, e.g., when the firm size is smaller and analysts' forecast dispersion is high.

\section{Time Series Model}
\label{sec:tsm}

\begin{figure}[htb]
	\centering
	\begin{tikzpicture}[xscale=1.1,yscale=0.5]
	\draw (-3.5,0) -- (4.5,0);
	\draw (-3.5,1) -- (4.5,1);
	\draw[] (-3.3,0.5) node{$\dots$};
	\draw[]                    (-3,0) rectangle node{$\predData_{t-3}$} ++(1,1);
	\draw[fill=green!30!cNorTb] (-2,0) rectangle ++(1,1) node[pos=.5] {$\predData_{t-2}$};
	\draw[fill=green!30!cNorTb]  (-1,0) rectangle ++(1,1) node[pos=.5] {$\predData_{t-1}$};
	\draw[fill=green!50!cNorTb] ( 0,0) rectangle ++(1,1) node[pos=.5] {$\predData_{t}$};
	\draw[fill=blue!50!cNorTb]                    ( 1,0) rectangle ++(1,1) node[pos=.5] {$\predData_{t+1}$};
	\draw[]  ( 2,0) rectangle ++(1,1) node[pos=.5] {$\predData_{t+2}$};
	\draw[]                    ( 3,0) rectangle ++(1,1) node[pos=.5] {$\predData_{t+3}$};
	\draw[] (4.3,0.5) node{$\dots$};
	
	\draw[->] (-3.5,-.4) -- (4.5,-.4) node[below, xshift=-2mm]{$t$};
	
	\draw (0.5,-1.4) -- (0.5,-1.0);
	\draw (1.5,-1.4) -- (1.5,-1.0);
	\draw[latex-latex] (0.5,-1.2) -- node[below]{$\predHorizont = 1$} (1.5,-1.2);
	
	\draw (-2,-1) -- (-2,-.6);
	\draw ( 1,-1) -- ( 1,-.6);
	\draw[latex-latex] (-2,-.8) -- node[below]{$\predFeaturewindow = 3$} (1,-.8);
	
	\draw [decorate,decoration={brace,amplitude=7pt}]
	(-2,1.1) -- (.95,1.1) node(A)[cNorTf,midway,above=2mm] {\footnotesize $x$};
	
	\draw [decorate,decoration={brace,amplitude=7pt}]
	(1.05,1.1) -- (2,1.1) node(B)[cNorTf,midway,above=2mm] {\footnotesize $y$};
	
	\draw[-latex] ([yshift=-.1] A.east) to [bend left=20] node[above=-1pt]{$\predAbbildung$} ([yshift=-.1] B.west);
\end{tikzpicture}
	\caption{
		Illustration of time series model for prediction of earnings of a company with quarterly reports $q_t$ at time step $t$.
		We seek a mapping \predAbbildung from pattern $x$ of earning data of the past to label $y$ of the predicted earning for the future $t=t+ \predHorizont$.
		The window size \predFeaturewindow describes the time span of considered past earnings.
	}
	\label{fig: time series prediction window}
\end{figure}
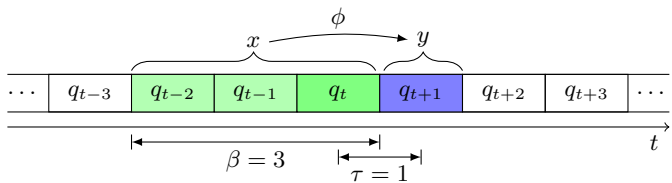

The goal in data-driven prediction based on time series is to find a function \predAbbildung that yields a future value $y$ based on the data of the past \predFeaturewindow time steps $x = (q_{t-\predFeaturewindow+1}, \dots, q_{t})$ (\cref{fig: time series prediction window}).
In this paper, the time-span between two time steps is 3 months.
A non-perfect predictor $\hat{\predAbbildung}(x) = \hat{y}$ can be evaluated using the mean squared error (\MSE) to the real value $y$.

To evaluate our model we compare it with the persistent model and the analysts forecast.
The persistence model is a simple baseline that uses the current value as a prediction for the next time step.
For each model the \MSE is calculated.
Therefore, larger deviations are more punished than smaller ones.

Since the difficulty of forecasting the given data varies greatly over time and between different companies, the error value in itself is not meaningful.
Therefore we use a relative comparison between the different models, namely the skill score (SS) \cite{Roebber1998Regime}:
\begin{equation}
	\text{SS}_{\text{\MSE}} = 1 - \frac{\text{\MSE}(m)}{\text{\MSE}(\text{base})},
\end{equation}
where $\text{\MSE}(m)$ is \MSE of the own model $m$ (\LSTM, \TCN) and $\text{\MSE}(\text{base})$ is the \MSE of the comparison model: persistent model%
\footnote{
	For the comparisons only data points are used for which analyst forecasts exist.
}
$pa$ or analyst forecast $a$.
The model under consideration is better (worse) than the reference model if the skill score is greater (less) than 0 \cite{Roebber1998Regime}.

\section{Data Preprocessing}
\label{sec:dp}

As input data, we use accounting data (e.g., total assets and cost of goods sold) from \computstatquarterly as well as daily stock market price and return data from \capsacronym{CRSP} (\dailyshares) as these are the most commonly used databases in accounting and finance research.
At first both datasets \computstatquarterly and \dailyshares are reduced to the most important parameters\footnote{%
	The following parameters of the data records are used.
	The parameters in brackets are only used for the assignment and selection of the samples.
	\computstatquarterly: (\dsparam{cusip}, \dsparam{fpedats}, \dsparam{ffi5}, \dsparam{ffi10}, \dsparam{ffi12}, \dsparam{ffi48}, \dsparam{financialfirm}, \dsparam{EPS\_Mean\_Analyst}), \dsparam{rdq}, \dsparam{epsfiq}, \dsparam{atq}, \dsparam{revtq}, \dsparam{nopiq}, \dsparam{xoprq}, \dsparam{apq}, \dsparam{gdwlq}, \dsparam{rectq}, \dsparam{xrdq}, \dsparam{cogsq}, \dsparam{rcpq}, \dsparam{ceqq}, \dsparam{niq}, \dsparam{oiadpq}, \dsparam{oibdpq}, \dsparam{dpq}, \dsparam{ppentq}, \dsparam{piq}, \dsparam{txtq}, \dsparam{gdwlq}, \dsparam{xrdq}, \dsparam{rcpq}
	\dailyshares: (\dsparam{cusip}, \dsparam{date}), \dsparam{ret}, \dsparam{prc}, \dsparam{vol}, \dsparam{shrout}, \dsparam{vwretd}
	\label{fn:selected parameters}
}
per time-step and firm.
Different value ranges of individual parameters $x$ are \enquote{normalized} and scaled using the total assets $\dsparam{atq}$: 
\begin{equation}
x' = \frac{x}{\max\{1, \dsparam{atq}\}} 
\end{equation}
and studentized:
\begin{equation}
z_i' = \frac{z_i - \overline{z}}{\sqrt{\frac{1}{n} \sum_i (z_i - \overline{z})^2}},
\end{equation}
where $\overline{z}$ is the mean of $z_i$.
Outliers of \dsparam{eps} which are partially erroneous are removed by using the first (last) percentile as minimum (maximum).
We create company samples of a given window size (number of quarters).
Smaller data gaps a filled using linear interpolation, while samples with larger gaps are rejected.
The quarterly data are merged with the corresponding daily stock data \dailyshares, which are also being studentized.

\section{Deep Neural Networks}
\label{sec:deep}

\begin{figure}[htb]
    \begin{minipage}[b]{.54\textwidth}
    	\centering
    	\includegraphics[width=\textwidth]{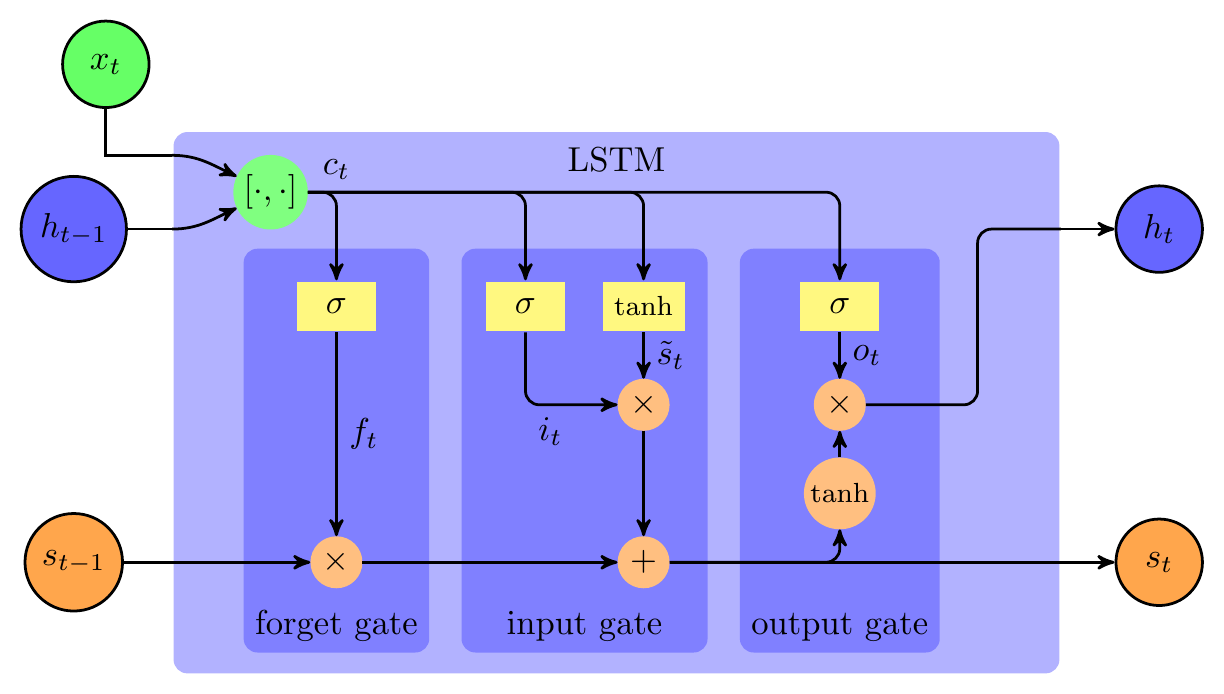}
    	\caption[\LSTM cell]{%
            Illustration of	\LSTM cell
    	}
    	\label{fig:lstm}
    \end{minipage}
    \hfill
    \begin{minipage}[b]{.44\textwidth}
    	\centering
    	\includegraphics[width=\textwidth]{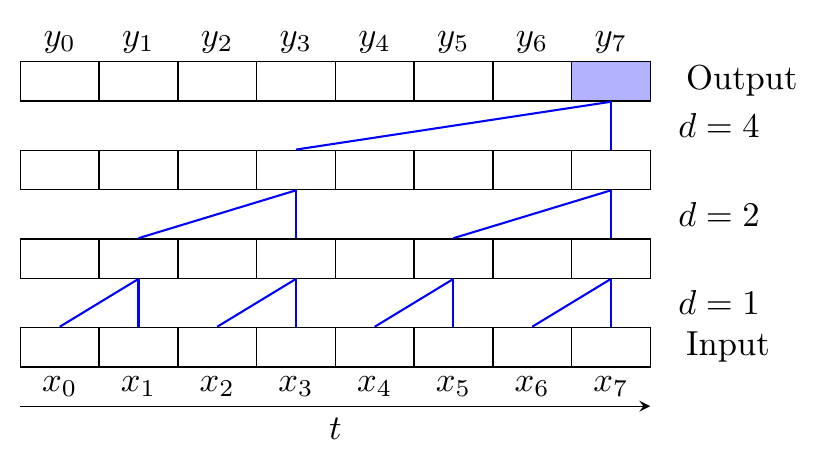}
    	\caption[Dilated causal convolution]{%
            Dilated causal convolution
    	}
    	\label{fig:dilated causal convolution}
    \end{minipage}
\end{figure}

An \LSTM network \cite{Hochreiter1997Long} belongs to the family of recurrent neural networks. It employs backward connections, which allow saving information in time.
\LSTM cells internally consists of three gates: forget, input, and output gates, see \cref{fig:lstm}.
An \LSTM cell employs internal states $h$ and $s$ propagated through time.
Yellow boxes represent ANN layers, orange circles represent element wise operations.
Input $x_t$ is concatenated with $h_{t-1}$ and fed to the forget, input, and output gates.
The forget gate determines which information should be forgotten, the input gate specifies to which amount the new input data is taken into account, and the output gate state specifies the information to output based on the internal state.
With these functional components, an \LSTM is well suited for time series data. \LSTM networks have successfully been applied to numerous domains, e.g., for wind power prediction \cite{Woon2017SpatioTemporal} and for speech recognition \cite{Graves2014Endtoend}.

A \TCN \cite{Bai2018Empirical} is a special kind of convolutional neural network \cite{LeCun1989Backpropagation}. While convolutional neural networks are primarily used for classification tasks in image, text or speech, \TCNs can be applied to time series data. \TCNs extend their counterparts by causal convolutions and dilated convolutions.
The \TCN has a one-dimensional time series input.
Causal convolutions only use the current and past information for each filter.
The dilation defines the distance between the used input data elements of each filter.
An example for both concepts is visualized in \cref{fig:dilated causal convolution} with a dilated causal convolution with kernel size $k = 2$ and dilations 1, 2, 4.
In our experiments we increase $d$ exponentially, i.e. $d_i=2^i$ and select an appropriate number of layers to cover the given time span.
\TCNs also find numerous applications, e.g., in satellite image time series classification \cite{Pelletier2019Temporal}.

\section{Experimental Analysis}
\label{sec:exp}

For our experiments we employed two datasets: \textbf{A} for the choice of a proper architecture and parameters and \textbf{B} for the final experiment with the selected best architecture.
The training set of A includes all samples whose predicted \EPS values lie in the period 2012 to end of 2016.
The last 10\% of the training set is used for validation only.
The test set is in the following half year after the training, so it is independent and has no unfair knowledge.
For data set B, the period is extended by half a year, so that its test data have not been seen before.

Each model is trained with a batch size 1024 for 1000 epochs and a dropout rate of 0.3 for each intermediate layer and the recurrent edges of an \LSTM layer.
Dense layers apply tanh as activation function, except for the last layer using a linear one.
The window size of \computstatquarterly and \dailyshares is set to 20, i.e., the last 20 quarters of earning reports and the last 20 daily stock market returns form a pattern.
The model is optimized using Adam and \MSE as loss.
Each epoch's best model w.r.t. validation error is used for testing.
Each experiment is repeated five times.
Statistics include mean and standard deviation.

\begin{figure}
    \centering
	\tikzset{
		every node/.style={
			draw,
			font=\scriptsize,
			minimum height=5mm,
		},
		node distance=3mm,
		>=stealth',
		inputq/.style={fill=green!30},
		inputs/.style={fill=green!30},
		lstm/.style={fill=red!30},
		tcn/.style={fill=orange!30},
		dense/.style={fill=blue!30},
		merge/.style={fill=cyan!30},
	}
	\subfloat[\label{fig:LSTM architecture}LSTM architecture]{%
		\begin{tikzpicture}
			\node[inputq] (inq) {quarters (20,19)};
			\node[lstm, right=of inq] (L1) {LSTM (20,76)};
			\node[lstm, right=of L1] (L2) {LSTM (38)};
			
			\node[dense, anchor=north east] (D3) at ([yshift=-3mm] L2.south east) {D (220)};
			\node[dense, left=of D3] (D2) {D (440)};
			\node[dense, left=of D2] (D1) {D (660)};
			\node[inputs, left=of D1] (ins) {shares (220)};
			
			\node[merge, anchor=west] (M) at ([xshift=3mm] $(L2.east)!.5!(D3.east)$) {merge};
			\node[dense, right=of M] (D4) {D (19)};
			\node[dense, right=of D4] (D5) {D (8)};
			\node[dense, right=of D5] (D6) {D (1)};
			
			\draw[->] (inq) -- (L1);
			\draw[->] (L1) -- (L2);
			\draw[->] (L2) -- (M);
		
			\draw[->] (ins) -- (D1);
			\draw[->] (D1) -- (D2);
			\draw[->] (D2) -- (D3);
			\draw[->] (D3) -- (M);
			
			\draw[->] (M) -- (D4);
			\draw[->] (D4) -- (D5);
			\draw[->] (D5) -- (D6);
		\end{tikzpicture}
	}
	
	\subfloat[\label{fig:TCN architecture}TCN architecture]{%
		\begin{tikzpicture}
			\node[inputq] (inq) {quarters (20,19)};
			\node[tcn, right=of inq] (T1) {TCN (f=32, k=3)};
			\node[dense, right=of T1] (Dx) {D (38)};
			
			\node[dense, anchor=north east] (D3) at ([yshift=-3mm] Dx.south east) {D (220)};
			\node[dense, left=of D3] (D2) {D (440)};
			\node[dense, left=of D2] (D1) {D (660)};
			\node[inputs, left=of D1] (ins) {shares (220)};
			
			\node[merge, anchor=west] (M) at ([xshift=3mm] $(Dx.east)!.5!(D3.east)$) {merge};
			\node[dense, right=of M] (D4) {D (19)};
			\node[dense, right=of D4] (D5) {D (8)};
			\node[dense, right=of D5] (D6) {D (1)};
			
			\draw[->] (inq) -- (T1);
			\draw[->] (T1) -- (Dx);
			\draw[->] (Dx) -- (M);
		
			\draw[->] (ins) -- (D1);
			\draw[->] (D1) -- (D2);
			\draw[->] (D2) -- (D3);
			\draw[->] (D3) -- (M);
			
			\draw[->] (M) -- (D4);
			\draw[->] (D4) -- (D5);
			\draw[->] (D5) -- (D6);
		\end{tikzpicture}
	}
	
	\caption{Visualization of selected \LSTM and TCN architectures.}
	\label{fig:selected architectures}
\end{figure}
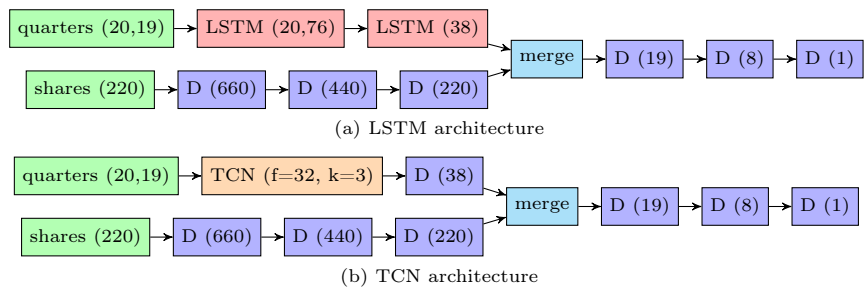

Furthermore, we have experimentally selected the best architectures as representatives for \LSTM and \TCN (\cref{fig:selected architectures}).
\computstatquarterly and \dailyshares are used as input (green).
The dimensions are given in parentheses.
Since the shares data is put into a dense layer (D), the time input $20 \times 11$ is flattened to 220.
After a few layers the two inputs are joined by a merge layer.
For the \TCN 32 filters and a kernel size of 3 were used.
The last dense layer with only one neuron outputs the predicted \EPS value.

As financial and non-financial companies show a significantly different behavior in many regards, we analyze the prediction in independent experiments.
\Cref{tab:financialfirms} compares the prediction performance with three different sets of companies: all companies (all), no financial companies (nofin), only financial companies (onlyfin). 
The data sets without financial firms usually give the best results.
The worst results are achieved when only financial companies are taken into account.

\begin{table}[htb]
	\scriptsize
	\centering
	\caption{%
		Selected architectures and parameters for three groups of companies: financial (onlyfin), non-financial (nofin), and all.
	}
	\label{tab:financialfirms}
\begin{tabular}{ll@{\hspace{3mm}}*{2}{r@{$\pm$}l}}
	\toprule
	 &  & \multicolumn{4}{c}{$\text{SS}_{\text{MSE}}$}\\
	\cmidrule(lr){3-6}
	type & comp & \multicolumn{2}{c}{$(m,pa)$} & \multicolumn{2}{c}{$(m,a)$}\\
	\midrule
	LSTM & all & \ltfbetter{0.466} & 0.030 & \ltfbetter{0.119} & 0.050\\
	 & nofin & \ltfbetter{\ltfmax{0.543}} & 0.013 & \ltfbetter{\ltfmax{0.146}} & 0.024\\
	 & onlyfin & \ltfbetter{\ltfmin{0.378}} & 0.025 & \ltfbetter{\ltfmin{0.100}} & 0.036\\
	\cmidrule{1-6}
	TCN & all & \ltfbetter{0.355} & 0.058 & -0.065 & 0.096\\
	 & nofin & \ltfbetter{\ltfmax{0.547}} & 0.015 & \ltfbetter{\ltfmax{0.154}} & 0.028\\
	 & onlyfin & \ltfmin{-0.001} & 0.155 & \ltfworse{\ltfmin{-0.450}} & 0.225\\
	\bottomrule
\end{tabular}
\end{table}

We test the best model an an independent dataset B. \Cref{tab:finref_b4_05} shows the results of the bests configurations of \cref{tab:financialfirms}.
The results for the non-financial companies are similar to the results observed before with an \MSE that is 12--13\% better than the analysts' predictions.
The predictions for all companies are slightly better, but worse than on dataset A.

\begin{table}[htb]
	\scriptsize
	\centering
	\caption{%
		Results on dataset B of optimal architectures and parameters grouped by financial sector affiliation.
	}
	\label{tab:finref_b4_05}
\begin{tabular}{ll@{\hspace{3mm}}*{2}{r@{$\pm$}l}}
	\toprule
	 &  & \multicolumn{4}{c}{$\text{SS}_{\text{MSE}}$}\\
	\cmidrule(lr){3-6}
	type & comp & \multicolumn{2}{c}{$(m,pa)$} & \multicolumn{2}{c}{$(m,a)$}\\
	\midrule
	LSTM & nofin & \ltfbetter{\ltfmax{\ltfmin{0.300}}} & 0.012 & \ltfbetter{\ltfmax{\ltfmin{0.122}}} & 0.015\\
	\cmidrule{1-6}
	TCN & nofin & \ltfbetter{\ltfmax{\ltfmin{0.308}}} & 0.014 & \ltfbetter{\ltfmax{\ltfmin{0.132}}} & 0.018\\
	\bottomrule
\end{tabular}
\end{table}

These results suggest that \LSTM networks and \TCNs are indeed able to provide meaningful earnings predictions.
Even after acknowledging for the variation across the repetitions (e.g., standard errors based on three repetitions), the range of significance (e.g., mean estimate plus/minus standard error) is well above zero in all cases.
This is remarkable, as we only used widely available public data on companies such as balance sheet information and stock market price and return data.
Hence, we can conclude that our networks outperform both the persistent model and the mean forecast of financial analysts based on a subsample of non-financial firms (e.g., manufacturing firms).

\section{Conclusion}
\label{sec:cons}

Our experimental analysis has shown that \LSTM networks and \TCNs are powerful models in the application of earnings prediction.
We base our prediction models on quarterly accounting data such as cost of goods sold and total assets as well as stock market price and return data.
Using these widely available time series data, the persistent model was significantly outperformed. 
The \LSTMs performed slightly better in our analysis using the same set of variables.
In the future, we will extend the experimental analysis to further data sets and integrate further domain knowledge to improve the financial predictions.
Our findings are relevant to both broker firms and investors.
Broker firms may want to consider developing \LSTM networks and \TCN to supplement their analysts' forecast.
Investors could build up their own forecast models using artificial intelligence, particularly when there are no forecasts available from financial analysts, which became a more urgent issue recently due to the drop in analyst coverage induced by regulation.

\urlstyle{rm}
\def\doi#1{\href{https://doi.org/#1}{\url{https://doi.org/#1}}}
\bibliographystyle{splncs04}
\bibliography{literature}
\end{document}